%% file: cooperative_queue_arch_v1.tex
\documentclass[conference,12pt,draftcls,onecolumn]{IEEEtran}
\IEEEoverridecommandlockouts
\usepackage{cite}
\usepackage[dvips]{graphicx}
\usepackage{color}
\usepackage{amsmath}
\usepackage{amssymb}

\newcommand{\beq}{\begin{eqnarray}}
\newcommand{\eeq}{\end{eqnarray}}

\newcommand{\cF}{\mathcal{F}}
\newcommand{\bbZ}{\mathbb{Z}}
\newcommand{\bff}{\mathbf{f}}
\newcommand{\bg}{\mathbf{g}}
\newcommand{\br}{\mathbf{r}}
\newcommand{\cM}{{\mathcal M}}
\newcommand{\T}{\mathcal{T}}
\newcommand{\cI}{\mathcal{I}}
\newcommand{\bone}{\mathbf{1}}
\newcommand{\bzero}{\mathbf{0}}
\newcommand{\bQ}{\mathbf{Q}}
\newcommand{\blam}{\boldsymbol\lambda}
\newcommand{\expec}{{\mathbb E}}
\newcommand{\bF}{\mathbf{F}}
\newcommand{\ba}{\mathbf{a}}
\newcommand{\bbb}{\mathbf{b}}

\begin{document}

\newtheorem{theorem}{Theorem}
\newtheorem{lemma}[theorem]{Lemma}
\newtheorem{cor}[theorem]{Corollary}
\newtheorem{remark}{Remark}
\newtheorem{result}{Result}
\newtheorem{defn}{Definition}
\newtheorem{exm}{Example}

\title{Queue-Architecture and Stability Analysis \\in Cooperative Relay Networks}

\author{\IEEEauthorblockN{Jubin Jose and Sriram Vishwanath}\\
\IEEEauthorblockA{Dept. of Electrical and Computer Engineering\\ University of Texas at Austin\\
Email: \{jubin, sriram\}@austin.utexas.edu}}
\maketitle

\begin{abstract}
An abstraction of the physical layer coding using bit pipes that are coupled through data-rates is insufficient to capture notions such as node cooperation in cooperative relay networks. Consequently, network-stability analyses based on such abstractions are valid for non-cooperative schemes alone and meaningless for cooperative schemes. Motivated from this, this paper develops a framework that brings the information-theoretic coding scheme together with network-stability analysis. This framework does not constrain the system to any particular achievable scheme, i.e., the relays can use any cooperative coding strategy of its choice, be it amplify/compress/quantize or any alter-and-forward scheme. The paper focuses on the scenario when coherence duration is of the same order of the packet/codeword duration, the channel distribution is unknown and the fading state is only known causally. The main contributions of this paper are two-fold: first, it develops a low-complexity queue-architecture to enable stable operation of cooperative relay networks, and, second, it establishes the throughput optimality of a simple network algorithm that utilizes this queue-architecture.
\end{abstract}

\section{Introduction}

Cooperative relaying is traditionally seen as a physical layer scheme for analyzing and designing wireless link layer protocols \cite{Sendonaris2003}, with limited network-layer insights originating from such schemes. Indeed, the not-so-uncommon perception is: whatever be the physical layer transmission/coding scheme, the network can abstract it into a ``rate region'' and then determine algorithms to stabilize queues, perform rate control and other tasks at the higher layers. From this perspective, it seems unimportant for researchers at either layer to learn much about the intricacies of the other.

There is a significant and growing body of work suggesting that such abstractions may not be accurate \cite{Dong2008} and that physical layer parameters must be included into the analysis. A large class of this work is based on signal-to-noise ratio (SNR) or signal-to-interference-and-noise ratio (SINR) models for the physical medium. While S(I)NR is a worthwhile abstraction for physical-layer schemes that ``treat interference as noise'', it is often overused and does not capture more involved physical layer transmission schemes \cite{Ephremides2010}. From information theory, it is well known that ``treating interference as noise'' represents a very limited class of transmission schemes, and a much larger class of schemes exist that achieve significantly higher throughput. Therefore, a framework that brings the information-theoretic coding scheme together with network-stability analysis is needed, to bridge the gap caused by the ``unconsummated union'' \cite{Ephremides1998}. In this paper, we explore building this bridge in the context of cooperative relay networks.

We emphasize that a natural separation between network stability and physical layer coding exists only for specific classes of networks (such as capacitated networks \cite{Bodas2007}) and not in general, and a joint framework is needed that can capture notions such as physical layer cooperation. In this paper, we focus on cooperative relay networks, where multiple reasons exist for combining network and physical layer aspects. 
\begin{itemize}
\item First, the rate-maximizing physical-layer coding strategy automatically imposes scheduling restrictions on the relays/transmitters in the network. For coherent combination at the receivers to be at all possible, all nodes involved must transmit simultaneously in that block. 
\item Second, it is codebooks and functions of codebooks being received, stored and transmitted by nodes and not traditional data packets. 
\item Finally, the codebook chosen by the source(s) determines the rate of transmission, which may or may not be alterable at intermediate nodes (this is a key distinction between general information-theoretic coding theorems and say, packetized or linear network coded systems where rate can always be varied at every node). For example, if a relay were to use amplify-and-forward or compress-and-forward as its physical-layer strategies, it has no control over rate and has a real vector as its ``packet''. 
\end{itemize}

Given the need for a joint physical and network layer framework for cooperative networks, the rest of the paper is organized as follows: in the next section, we present a brief summary of cooperative relay networking from a physical layer perspective. In Section \ref{sec:mainresults}, we present our main results in this paper. In Section \ref{sec:sysmodel}, we describe our system model in the context of heterogeneous cellular networks. In Section \ref{sec:rate}, we describe cooperative schemes for such networks in detail and present a queue-architecture that enables both efficient and optimal operation of the network. In Section \ref{sec:main}, we present the main algorithm for operating such networks, and establish that this algorithm is throughput-optimal. We conclude with Section \ref{sec:conclude}.

\section{Background: Cooperative  Relay Networks}
\label{sec:cooprelay}

Cooperative relay networks have been researched extensively since the ``MIMO effect'' was established. Until recently, it was considered hard if not impractical for nodes to coordinate transmissions to enable cooperative relaying. However, emerging heterogeneous cellular networks are increasingly moving in the direction of standardizing and evaluating schemes with node cooperation \cite{3GPP,Sawahashi2010}. As cell sizes decrease, an increase in cell edges and interference requires node cooperation to increase throughput, and cooperative relaying is an important step in making this happen.

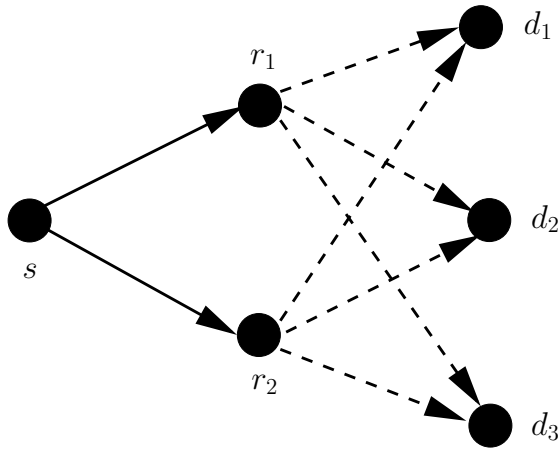
\begin{figure}
\centering
\scalebox{0.6}{\input{./two_hop_network.pstex_t}}
\caption{Two-hop Cooperative Network}
\label{fig:two_hop}
\end{figure}

Figure \ref{fig:two_hop} shows the most basic configuration that incorporates cooperative relaying in heterogeneous cellular networks. To motivate this setting, we take the example of a macrocellular network. Here, the source node $s$ corresponds to the macro-cell base-station, the relay nodes $r_1$ and $r_2$ correspond to pico-cell base-stations and the destination nodes $d_1$, $d_2$ and $d_3$ correspond to mobiles. We focus on the downlink scenario where the source $s$ has independent messages/bits for the mobiles. The relays' role is to help the source in transmitting these messages. Further, we assume a half-duplex cooperative  constraint so that either the first-hop or the second-hop links can be activated at any given time, with no direct-links from the source to the destinations. A more general and detailed system model for such cooperative relay networks is provided in Section \ref{sec:sysmodel}.

Even for such simple networks with two relays and one destination and fixed channels,  information-theoretic capacity is not yet known. However, there has been significant progress in developing cooperative communication schemes for such systems by using coherence and physical-layer coordination among nodes. There are multiple strategies studied in literature that enable this coordination, referred to as {\em forwarding} schemes. One such scheme of interest is the so-called decode-and-forward scheme that requires relays to decode messages. In contrast to traditional networks, the relays decode common messages, that are then transmitted cooperatively. However, the relays still have decoded messages or packets as in traditional networks. In \cite{J:YB07}, the authors develop a throughput-optimal network algorithm that can handle common messages. In \cite{C:YB07}, the authors consider more general network configurations, but the applicability is still limited to decode-and-forward schemes with fixed channels. In essence, all of these apply only in packet-in-packet-out networks. Complimentary to this is the work on optimal resource allocation for non-cooperative wireless networks \cite{J:TE92,Lin_Shroff_Srikant_06,B:GNT06} (and references therein).

In our effort, we do not want to constrain the system to a packet-in-packet-out framework. We desire that the relays use {\em any} information-theoretic cooperative coding strategy of its choice, be it amplify/compress/quantize or any alter-and-forward scheme. This couples coding, resource allocation and stability into one joint problem, and the analyses in \cite{J:YB07,J:TE92,B:GNT06} and the vast literature on non-cooperative networks do not apply. Even the analyses in \cite{J:YB07,C:YB07} for decode-and-forward cooperative networks do not apply. This motivates the need for a new framework and stability analysis.

Before proceeding to describe our results, a note to state the obvious: if the channel state is fixed and thus its capacity is precomputed, a simple static split scheme will ensure stable operation while maximizing the information theoretic rate (region) for the network. The challenge, of course, is when the fading state distribution and input arrival rates are unknown, and the fading state can only be observed causally. For example, consider a fading channel with block fading of $T$ symbols each. When $T$ is much smaller compared to the packet duration (or equivalently the channel-coding duration), queueing/buffering of packets at relays is not required as the first-hop and second-hop can be operated sequentially without reducing data-rates. When $T$ is comparable to (or larger than) the packet duration, queueing of packets at relays can provide significant gains in terms of data-rates. Furthermore, when $T$ is roughly the same as the packet duration, queueing at relays is inevitable as the source does not know the fading state of the second-hop while encoding the packet. In this paper, we focus on the second scenario when $T$ is larger than the packet/codeword duration. Given that the channel distribution is unknown and the fading state is only known causally, we ask the question:  Is it possible to stabilize the network while operating it close to the boundary of its information-theoretic rate region?

\section{Main Results}
\label{sec:mainresults}

The answer to the preceding question  in Section \ref{sec:cooprelay} is ``yes", which is proved for a simpler network with two relays and one destination in \cite{Jose-Ying-Vishwanath-09}. In this setting, for cooperative schemes such as amplify/quantize-and-forward and partial-decode-and-forward, the relays receive and transmit real-valued ``packets''. In order to accomplish this in \cite{Jose-Ying-Vishwanath-09}, we introduce a new ``state-based'' virtual-queue-architecture for these real-valued ``packets'', and develop a throughput-optimal network algorithm that does not require the knowledge of the fading distribution. Each ``state'' corresponds to a vector comprised of the {\em entire} channel-state of each link in the network. This approach, although analytically very helpful, suffers from a major issue that makes it practically uninteresting -  requiring that a virtual-queue be maintained for each channel-state at each node in the network leads to an explosion of queues, even for simple network configurations. Moreover, the approach in \cite{Jose-Ying-Vishwanath-09} is particular to a single destination setting. In this paper, we develop a simpler queue-architecture to enable stable operation of cooperative relay networks. Further, we generalize it to any forwarding scheme with multiple destinations. 

The virtual-queue-architecture we introduce in this paper is primarily {\em encoding-based}. This architecture is motivated by the manner in which adaptive modulation and coding is currently implemented in practice. In systems today, the source node implements a limited number of encoding schemes (encoding functions and rate-vectors). Each encoding scheme is designed so that it can be successfully employed for a particular subset  of states. Even though encoding schemes belong to a finite (and usually small) set, the mapping functions at the relays and the decoding functions at the destinations are usually state-dependent. A queue-architecture that keeps virtual queues at the relays for each state corresponding to the first-hop and each encoding scheme is sufficient. This considerably reduces the number of virtual queues that must be maintained while still remaining a ``sufficient statistic", i.e., these encoding-based queues are a sufficiently rich representation for us to develop throughput optimal algorithms using them. 
Using this new and somewhat intuitive virtual-queue-architecture, we develop a network algorithm that has the following properties. 
\begin{enumerate}
\item It does not require the knowledge of the fading distribution. 
\item It does not require the knowledge of the arrival rates. 
\item It keeps all the queues stable for any arrival rate-vector within the throughput region, i.e., it is throughput-optimal. 
\end{enumerate}
Note that limiting ourselves to a small set of possible encoding schemes and rates inherently reduces the network's information-theoretic rate region. The more fine-grained the encoding schemes and resulting queue-architecture, the smaller the loss in rate region. However, note that the encoding-based queue-architecture itself does not introduce any sub-optimality.

In summary, we introduce and study a new encoding-based queue-architecture, which is inspired by an adaptive coded modulation system analyzed and implemented at the physical layer in systems today. However, in today's systems, there is limited interaction, if any, between network-layer algorithms and adaptive coding/modulation, and we argue that  coupling them together can be very useful in both the analysis and design of cooperative relay networks. Indeed, we show that such a queuing architecture can result in throughput optimal algorithms, and the network can achieve its information-theoretic rate region corresponding to its choice of encoding/decoding strategies while maintaining stability.




\section{System Model}
\label{sec:sysmodel}

We consider discrete-time two-hop cooperative networks that include the network shown in Figure \ref{fig:two_hop}. We allow for arbitrary number of relays and destinations, i.e, the network consists of a source node denoted by $s$, $N$ relay nodes denoted by $r_1, r_2, \ldots, r_N$, and $K$ destination nodes denoted by $d_1, d_2, \ldots, d_K$. The source has independent messages for all the destinations. The relays aid in transmitting these messages to their respective destinations. Throughout this paper, ``first-hop'' refers to the links from the source to the relays, and ``second-hop'' refers to the links from the relays to the destinations.  At any given time, half-duplex and cooperative-communication constraints require that either the first-hop or the second-hop can be activated and not both. The presence of direct links from source to destinations will not invalidate the analysis presented in this paper, but would render it considerably harder. For simplicity, we assume that they are absent and thus concentrate on equal-path length networks.

The channel model does not directly impact the queue-architecture, and thus the network algorithm and stability analysis presented in this paper. The channel is state dependent, and the joint-state distribution be unknown. A particular channel model of interest is a linear interaction model with additive white Gaussian noise (AWGN). In the context of an AWGN channel, an example of state is a multiplicative fading parameter. We focus on a framework with i.i.d. block-fading model with a block-length of $T$ symbols in the remainder of this paper. The channels remain constant for the duration of one block, and then change to a new (independent) realization from an underlying distribution from block to block. Let $t \in \bbZ_+$ denote the channel fading blocks, and let $\cF$ denote the fading state-space, which is assumed to be discrete. In block $t$, $\bff_1[t] \in \cF^N$ denotes the fading realization for the first-hop and $\bff_2[t] \in \cF^{NK}$ denotes the fading realization for the second-hop. The combined fading-state is denoted by $\bff[t]=(\bff_1[t],\bff_2[t])$. The corresponding random vectors are denoted by $\bF_1[t]$, $\bF_2[t]$ and $\bF[t]$. Note that $\bF[t]$ is i.i.d. over time, but can be spatially correlated. Let the probability that $\bF[t]$ takes value $\bff$ be $\pi_{\bff}$. This is the underlying probability distribution that is unknown to the central controller.

Next, we explain the time-scales in which network and channel parameters evolve in our system. The coherence time $T$ is assumed to be comparable to the channel-coding length in symbols. For the ease of presentation, the ``packet" (which is either the channel codeword or any real-vector representing the actual data packet) length is assumed to be equal to the coherence time $T$. It is straightforward to extend the analysis when the ``packet'' length is a sub-multiple of the coherence time $T$. Each ``packet" is transmitted on the first-hop and the second-hop exactly once. These transmissions need not happen in consecutive time-blocks, i.e., these ``packets" can be buffered at the relays. The coding performed at the source, the mappings performed at the relays, and the decoding at the destinations can be arbitrary, i.e., this includes any and all schemes that are information-theoretically capacity-optimal or, if capacity is unknown, then the best known coding scheme. Further, we assume that the instantaneous fading-state  is causally known globally to the central controller. In other words, prior to transmission, the central controller is aware of the entire network channel state for that particular time-block. 


\subsection{Notation}
Vectors are denoted by bold letters. For vectors, equality and inequality operators are defined component-wise.  $\ba \cdot \bbb$ denotes the dot product of $\ba$ and $\bbb$. $|\cdot|$ denotes the cardinality of a set. $\mathbf{1}_{\{E\}}$ denotes the indicator function of event $E$.  $(a)^+$ denotes $\max(a,0)$. $\expec[\cdot]$ denotes the expectation operator.

\section{Achievable Rates \& Queue-Architecture}
\label{sec:rate}

The notion of a ``packet'' here is different from traditional networks where a packet is decoded at all intermediate relays, and is usually meant for one destination. In this paper, the term ``packet'' refers to the set of coded symbols transmitted/received in the network. Note that  each of the relays receives a different noisy version of the transmitted vector (transmitted ``packet"),  which is subsequently mapped to a transmit vector (``packet") at each relay. Again, the destinations receive a noisy version of a linear combination of relays' transmit ``packets". In this paper, we refer to the physical-layer signalling vectors as {\em packets} at each node in the network. We choose to use this language as the entire network layer analysis is based on understanding the dynamics of these transmit vectors as they traverse the system. Consider a packet that is transmitted from the source to the $K$ destinations. Let this packet be transmitted on the first-hop during block $t_1$, and be transmitted on the second-hop during block $t_2$. Then, $\bg = (\bff_1[t_1],\bff_2[t_2])$ is said to be the ``state'' seen by this packet. Note that this notion of state is different from physical channel fading state, but  is it of equal importance in our analysis.

A packet transmitted by the source is received by all the destinations in two hops, but the amount of information each destination receives varies depending on the encoding rates. Given a state seen by the packet, the set of encoding rates that can be supported is known as the rate region for the given state. An extremely challenging problem even in the single destination setting is to find the set of all achievable rates, or the capacity region for the given state. Even though the capacity region is unknown in most cases, there are many efficient cooperative communication schemes that have been developed. Therefore, the main aims of this paper are: (\emph{i}) to develop a queue-architecture that can support existing (and future) cooperative schemes, and (\emph{ii}) to develop a throughput-optimal network algorithm using this queue-architecture.

The queue-architecture developed in \cite{Jose-Ying-Vishwanath-09} for single-destination setting keeps ``virtual'' queues at relays for every state. Suppose that each rate-region can be quantized such that the convex-hull of the set of quantized rate-vectors is ``nearly'' same as the rate-region itself. Further, let us assume that the rate corresponding to each destination have to be quantized to $L$ levels. Now, a direct extension of the state-based virtual-queue-architecture would require ``virtual'' queues at relays for each state and each quantized rate-vector, which results in $L^K|\cF|^{K(N+1)}$ ``virtual'' queues. This scales exponentially in the number of destinations $K$. Clearly, such a queue-architecture is not scalable in practice, and will face implementation issues.

In order to design a low-complexity queue-architecture, we exploit the fact that practical systems implement limited number of encoding schemes, as in the case of adaptive modulation and coding. For example, the source might choose to encode only two destinations at a time using superposition encoding. In this case, the total number of encoding schemes would be $K(K-1)L^2$. In another example, the source might choose to encode at limited boundary rate-vectors again with superposition encoding. Let $\cM$ denote the set of encoding schemes, and $\br_m$ denote the rate-vector corresponding to each encoding scheme $m \in \cM$. Given that $|\cM| \ll L^K|\cF|^{KN},$ a queue-architecture needs to support these limited choices. While a queue-architecture can take advantage of this, it needs to allow for arbitrary mapping at the relays and decoding at the destinations. These are usually state-dependent, for example, an amplify-and-forward mapping is state-dependent.

Before describing our queue-architecture, we characterize the throughput region of the two-hop cooperative network. For this, we assume the knowledge of the fading distribution. Define $\cI=\{(m,\bg)|m\in \cM \text{ can be supported by state }\bg \in \cF^{(N+1)K}\}$, which represents whether an encoding scheme is supported by a state or not\footnote{We do not explicitly deal with packet error rate, as it is assumed that the achievable rate-vector is defined appropriately with required packet error rate.}. Now, let $\bff=(\bff_1,\bff_2)$ be any fading-state where $\bff_1$ is the fading-state of first-hop and $\bff_2$ is the fading-state of second-hop. Similarly, let $\bg=(\bg_1,\bg_2)$ by any state. We define $\hat{\cF}=\cF^{(N+1)K}$, $\cI_1=\{(\bff,\bg)|\bg_{1}=\bff_1\}$, and $\cI_2=\{(\bff,\bg)|\bg_{2}=\bff_2\}$. With the above definitions, the throughput region of the network is characterized in the following lemma.

\begin{lemma}
\label{lem:thruput}
A rate-vector $\hat{\br}$ is in the throughput region denoted by $\T$ only if there exists $a_{\bff}^{m,\bg}\geq 0$ and $b_{\bff}^{m,\bg}\geq 0$ for all $m \in \cM$, $\bg \in \hat{\cF}$ and $\bff \in \hat{\cF}$ such that
\begin{eqnarray}
&&{\hat{\br}=\sum_{m, \bg, \bff} \left( \pi_{\bff}a_{\bff}^{m,\bg} \br_{m} \mathbf{1}_{\{(\bff,\bg)\in \cI_1\}}\mathbf{1}_{\{(m,\bg)\in \cI \}} \right),}\label{eq: fc1}\\
&&{\sum_{\bff \in \hat{\cF}} \pi_{\bff}a_{\bff}^{m,\bg} \mathbf{1}_{\{(\bff,\bg)\in \cI_1\}}=} \sum_{\bff \in \hat{\cF}} \pi_{\bff}b_{\bff}^{m,\bg} \mathbf{1}_{\{(\bff,\bg)\in \cI_2\}},   \forall (m,\bg)\in \cI\label{eq: fc2},\\
&&{\sum_{m, \bg}a_{\bff}^{m,\bg} +b_{\bff}^{m,\bg}  \leq 1, \forall \bff.}
\label{eq: overall}
\end{eqnarray}
\end{lemma}
\begin{IEEEproof}
Let $a_{\bff}^{m,\bg}$ be the fraction of time for which packets corresponding to encoding scheme $m$ and state $\bg$ is transmitted from the source to the relays when the system is in fading state $\bff.$ Similarly, let $b_{\bff}^{m,\bg}$ be the fraction of time for which these packets are transmitted from the relays to the destinations. (\ref{eq: fc1}) is flow conservation constraint for the source, and (\ref{eq: fc2}) is the flow conservation constraint for each encoding scheme and state. (\ref{eq: overall}) is the time conservation constraint for each fading-state. A central controller with the knowledge of the fading distribution can achieve these rates using static time-division.
\end{IEEEproof}

An immediate corollary of this lemma is the following.
\begin{cor}
The throughput region $\T$ is convex. 
\end{cor}

{\bf Encoding-based Queue-architecture:} At the source node $s$, there are $K$ queues consisting of bits (or data) corresponding to the $K$ destinations. We denote the queue at the source corresponding to $k$-th destination by $Q_s^k$ with queue-length $Q_s^k[t]$ during block $t$. There is an exogenous i.i.d. arrival process $A^k[t]$ of data-bits into $Q_s^k$ with mean rate $\lambda_k T$ bits/block and bounded variance. The vector of arrival rates $\lambda_k$ is denoted by $\blam$. At each relay (say $n$), we keep virtual queues corresponding to each encoding scheme $m$ and each fading state for the first-hop $\bg_1$ denoted $Q_n^{m,\bg_1}$ with queue-length $Q_n^{m,\bg_1}[t]$ during block $t$. This queue consists of real-valued packets encoded at rate $\br_m.$ Since we keep virtual queues for each fading state corresponding to the first-hop, the mapping function performed at the relays can a function of the fading state. Similarly, the decoding function can be a function of the fading state. With this queue-architecture, the number of virtual queues at each relay is $|\cM||\cF|^N.$ This is considerably less compared to the number of virtual queues required in the state-based approach, and thus provides a low-complexity queue-architecture. Note that the gain is high in the setting when the number of destinations are large and number of relays are small, which is the case in cellular systems. 

The queue dynamics is as follows: During block $t$, if the fading state for the first-hop is $\bg_1$ and if the central controller decides that the source should transmit a packet using encoding scheme $m$, then the following queues get updated:
\begin{eqnarray}
\label{eq:Qs1}
Q_s^k[t+1] &=& (Q_s^k[t] + A^k[t] - r_m^kT)^+,  \forall k,\\
\label{eq:Qn1}
Q_n^{m,\bg_1}[t+1] &=& Q_n^{m,\bg_1}[t] + T,  \forall n.
\end{eqnarray}
During block $t$, if the fading state for the second-hop is $\bg_2$, then the central controller can decide to transmit packets from queues $Q_n^{m,\bg_1},\forall n$ for some given $m$ and $\bg_1$ only if $(m,\bg_1,\bg_2)\in \cI $. This ensures that the packet is received successfully at all the destinations. In this case, the following queues get updated:
\begin{eqnarray}
\label{eq:Qs2}
Q_s^k[t+1] &=& Q_s^k[t] + A^k[t],  \forall k,\\
\label{eq:Qn2}
Q_n^{m,\bg_1}[t+1] &=& (Q_n^{m,\bg_1}[t] - T)^+,  \forall n.
\end{eqnarray}

Next, we address the question of designing a central controller that does not have the knowledge of the arrival rates or the fading state distribution. 

\section{Throughput-Optimal Network Algorithm}
\label{sec:main}

In this section, we show that a throughput-optimal central controller can be designed without the knowledge of the arrival rates or the fading state distribution. Since cooperative schemes require strong node coordination, the centralized nature of the algorithm does not create additional system requirements. The following algorithm is motivated from back-pressure based Max-Weight algorithms for non-cooperative networks. 

{\bf Back-pressure-based Algorithm:} In every block, the central controller makes decisions based on the current fading state of the system and the current queue-lengths. Let the fading-state during block $t$ be $\bff[t]=(\bff_1,\bff_2)$. The network algorithm run by the controller is as follows:

\begin{enumerate} 
\item It computes 
\begin{eqnarray}
A = \max \limits_{m} \sum_{k}\left(Q_s^k[t] - r_m^k \sum_{n=1}^NQ_{n}^{m,\bff_1}[t]\right)r_{m}^k \nonumber
\end{eqnarray}
and an optimal parameter $m^*$ for this problem. 
\item It computes
\begin{eqnarray}
B = &\max \limits_{m,\bg_1} &(\br_m \cdot \bone)^2 \sum_{n=1}^NQ_{n}^{m,\bg_1}[t],\nonumber \\
&\text{s.t.}& (m,(\bg_1,\bff_2))\in \cI, \nonumber
\end{eqnarray}
and a set of optimal parameters $\hat{m}$ and $\hat{\bg}_1$ for this problem. 
\item If $A\ge B$, then the central controller decides to transmit a packet from the source to the relays using encoding scheme $m^*$. 
\item Otherwise, the central controller decides to transmit a packet from queues $Q_n^{\hat{m},\hat{\bg}_1},\forall n$, i.e., from the relays to the destinations.
\end{enumerate}
The controller repeats steps $1-4$ in every block. 

The following theorem provides a strong theoretical guarantee on the throughput performance of this algorithm.

\begin{theorem}
\label{thm:main}
The above algorithm stochastically stabilizes all the queues for any $\blam$ if there exists $\epsilon>0$ such that $\blam+\epsilon\bone$ is within the throughput region given in Lemma \ref{lem:thruput}, i.e., the underlying network Markov chain is positive recurrent. In simple terms, the algorithm is throughput-optimal. 
\end{theorem}

Before proceeding to the proof of this theorem, we state the following lemma that is used in the proof.
\begin{lemma}
\label{lem:for_main}
Suppose that there exists $\epsilon>0$ such that $\blam+\epsilon\bone$ is within the throughput region. Then, there exists $a_{\bff}^{m,\bg} \ge 0$, $b_{\bff}^{m,\bg} \ge 0$ and $\delta > 0$ such that the following set of conditions are satisfied:
\begin{eqnarray}
&&\lambda_k - \sum_{m,\bg,\bff} (\pi_{\bff} r_{m}^{k} a_{\bff}^{m,\bg}) \le -\delta, \forall k, \nonumber \\
&&\sum_{\bff} \pi_{\bff} (a_{\bff}^{m,\bg}-b_{\bff}^{m,\bg}) \le -\delta, \forall m,\bg, \nonumber \\
&&\sum_{m,\bg}a_{\bff}^{m,\bg} +b_{\bff}^{m,\bg}  \leq 1, \forall \bff, \nonumber \\
&&a_{\bff}^{m,\bg} = 0,  \forall (\bff,\bg) \notin \cI_1,\forall (m,\bg) \notin \cI,\nonumber \\
&&b_{\bff}^{m,\bg} = 0,  \forall (\bff,\bg) \notin \cI_2,\forall (m,\bg) \notin \cI. \nonumber
\end{eqnarray}
\end{lemma}
\begin{IEEEproof}
The proof of this lemma is fairly straightforward, and is omitted for brevity.
\end{IEEEproof}

\subsection{Proof of Theorem \ref{thm:main}}
Since the queues form a Markov chain, we use Foster-Lyapunov theorem in order to prove the stability \cite{Meyn-Tweedie-93,B:A03}. Without loss of generality, we assume that $\br_{m} \ne \bzero,\forall m$. Otherwise, those queues at the relays can be removed without affecting the throughput region and the stability of the system. Now, consider the Lyapunov function
$$V(\bQ[t]) = \sum_{k}\left(Q^k_s[t]\right)^2 + \sum_{n=1}^{N}\sum_{m,\bg_1} \left(\br_m\cdot \bone Q_{n}^{m,\bg_1}[t]\right)^2,$$ 
where $\bQ[t]$ denotes the vector of all queue lengths.

Next, we consider an optimization problem that captures the algorithm given in this section. Consider a fading-state $\bff$ and the following discrete optimization problem:
\begin{eqnarray}
&\max \limits_{\alpha_{\bff}^{m,\bg} , \beta_{\bff}^{m,\bg}} & \sum_{m,\bg,k} \left[\left(Q_s^k[t] - r_m^k \sum_{n=1}^NQ_{n}^{m,\bg_1}[t]\right)r_{m}^k \alpha_{\bff}^{m,\bg}\right]  \nonumber \\
\label{back_pressure}
& & +\sum_{m,\bg} \left[(\br_m \cdot \bone)^2 \left(\sum_{n=1}^NQ_{n}^{m,\bg_1}[t]\right)\beta_{\bff}^{m,\bg}\right], \\
&\text{s.t.} & \sum_{m,\bg} (\alpha_{\bff}^{m,\bg} + \beta_{\bff}^{m,\bg}) \le 1, \nonumber \\
& & \alpha_{\bff}^{m,\bg} = 0,  \forall (\bff,\bg) \notin \cI_1, \nonumber \\
& & \beta_{\bff}^{m,\bg} = 0,  \forall (\bff,\bg) \notin \cI_2,\forall (m,\bg) \notin \cI, \nonumber \\
& & \alpha_{\bff}^{m,\bg}, \beta_{\bff}^{m,\bg} \in \{0,1\},  \forall m,\bg. \nonumber
\end{eqnarray}
It is fairly straightforward to check that the algorithm given in this section results from this optimization problem. We remark that this optimization has many redundant variables that are introduced for the purpose of the proof.

Let an optimal assignment to the optimization problem in (\ref{back_pressure}) be ${\hat{\alpha}}_{\bff}^{m,\bg},{\hat{\beta}}_{\bff}^{m,\bg}$. Now, from (\ref{eq:Qs1}), (\ref{eq:Qs2}), (\ref{eq:Qn1}) and (\ref{eq:Qn2}), we can bound queue-lenths during block $t+1$ as follows:
\begin{eqnarray}
(Q_s^k[t+1])^2 &= &\left(Q_s^k[t] + A^k[t] - \left(\sum_{m,\bg}r_{m}^kT{\hat{\alpha}}_{\bff}^{m,\bg}\right)\right)^2 \nonumber \\
&\le &(Q^k_s[t])^2 + ({A}^k[t])^2 + \left(\sum_{m,\bg}r_{m}^kT{\hat{\alpha}}_{\bff}^{m,\bg}\right)^2 \nonumber \\&&- 2Q^k_s[t]\left(\sum_{m,\bg}r_{m}^kT{\hat{\alpha}}_{\bff}^{m,\bg}-{A}^k[t]\right), \forall k,\nonumber
\end{eqnarray}
\begin{eqnarray}
{(\br_m\cdot \bone Q_{n}^{m,\bg_1}[t+1])^2} &\le&\left(\br_m\cdot \bone Q_{n}^{m,\bg_1}[t] + \br_m\cdot \bone T \sum_{\bg_2}\left({\hat{\alpha}}_{\bff}^{m,\bg} - {\hat{\beta}}_{\bff}^{m,\bg}\right)\right)^2 \nonumber \\
&= &\left(\br_m\cdot \bone Q_{n}^{m,\bg_1}[t]\right)^2 + \left(\br_m\cdot \bone T \sum_{\bg_2}\left({\hat{\alpha}}_{\bff}^{m,\bg} - {\hat{\beta}}_{\bff}^{m,\bg}\right)\right)^2 \nonumber \\ 
&&-2(\br_m\cdot \bone)^2 Q_{n}^{m,\bg_1}[t] T \sum_{\bg_2}\left({\hat{\alpha}}_{\bff}^{m,\bg} - {\hat{\beta}}_{\bff}^{m,\bg}\right),\forall m, \bg_1. \nonumber
\end{eqnarray}

Applying the law of iterated expectations, we obtain
\begin{eqnarray}
{\mathbf{E}\left[V\left(\bQ\left[t+1\right]\right)-V\left(\bQ\left[t\right]\right)|\bQ\left[t\right]\right]-M}& \le& \sum_{\bff} \pi_{\bff} \left[- \sum_{k}2Q^k_s[t]\left(\sum_{m,\bg}r_{m}^kT{\hat{\alpha}}_{\bff}^{m,\bg}-\lambda_kT\right)- \right.\nonumber \\ 
 & &\left. \sum_{m,\bg_1,n} \left(2(\br_m\cdot \bone)^2 Q_{n}^{m,\bg_1}[t] T \sum_{\bg_2}\left({\hat{\alpha}}_{\bff}^{m,\bg} - {\hat{\beta}}_{\bff}^{m,\bg}\right)\right)\right] \nonumber \\
 \label{eq:before_lp_relax}
 & =&2T\left[\sum_{k}Q_s^k[t]\left(\lambda_k - \sum_{m,\bg,\bff} \left(\pi_{\bff} r_{m}^{k} {\hat{\alpha}}_{\bff}^{m,\bg}\right)\right)+ \right.\nonumber \\ 
 &&\left.\sum_{m,\bg,n} (\br_m\cdot \bone)^2 Q_n^{m,\bg_1}[t] \left(\sum_{\bff} \pi_{\bff} \right)\right].
\end{eqnarray}
where $M$ is a finite positive value, as the variance associated with the arrival processes are bounded and the throughput region is compact.

Let $a_{\bff}^{m,\bg},b_{\bff}^{m,\bg}$ be the values given by Lemma \ref{lem:for_main}. Now, substituting values $a_{\bff}^{m,\bg}$ instead of $\hat{\alpha}_{\bff}^{m,\bg}$ and $b_{\bff}^{m,\bg}$ instead of $\hat{\beta}_{\bff}^{m,\bg}$ in right hand side of (\ref{eq:before_lp_relax}) increases its value. This is due to the following reason. First, consider the linear program (LP) obtained by relaxing the integer constraints of the optimization problem (\ref{back_pressure}) and introducing non-negativity constraints. This relaxation is tight as LPs have at least one optimal solution which is a boundary point. Next, the possible values for $a_{\bff}^{m,\bg},b_{\bff}^{m,\bg}$ is a subset of the feasible set for the LP. Therefore, by substituting results from Lemma \ref{lem:for_main} in (\ref{eq:before_lp_relax}), we have
\begin{eqnarray}
{\mathbf{E}\left[V(\bQ[t+1])-V(\bQ[t])|\bQ[t]\right]-M} &\le &2T\left[\sum_{k}Q_s^k[t]\left(\lambda_k - \sum_{m,\bg,\bff} (\pi_{\bff} r_{m}^{k} a_{\bff}^{m,\bg})\right)+ \right.\nonumber \\ 
 &&\left.\sum_{m,\bg,n} (\br_m\cdot \bone)^2 Q_n^{m,\bg_1}[t] \left(\sum_{\bff} \pi_{\bff} (a_{\bff}^{m,\bg}-b_{\bff}^{m,\bg})\right)\right]\nonumber \\
 \label{eq:negdrift}
 &\le &-2T\delta\left[\sum_{k}Q_s^k[t]+ \sum_{m,\bg,n} (\br_m\cdot \bone)^2 Q_n^{m,\bg_1}[t] \right].
\end{eqnarray}
Now, from (\ref{eq:negdrift}), it is fairly straightforward to see that there is strict negative drift except on a compact subset of the set of queue-lengths. This completes the proof.
\QED

\section{Conclusion}
\label{sec:conclude}
In this paper, we develop encoding-based queue architecture for cooperative relay networks. Cooperative relay networks are fundamentally different from traditional capacitated and non-cooperative wireless networks as they require physical layer coordination. This physical layer coordination cannot be abstracted out at the network layer in terms of bits-in-bits-out models, and thus a stability analysis that incorporates both the physical layer encoding and the network layer dynamics is needed, as performed in this paper. The encoding-based queue architecture is a succinct representation needed for generating network stabilizing algorithms. Using this queue-architecture, we show that throughput-optimal network algorithms can be developed even when the fade-distribution and input queue distributions are unknown.


\end{document}

%% file: two_hop_network.pstex_t
\begin{picture}(0,0)%
\includegraphics{two_hop_network.pstex}%
\end{picture}%
\setlength{\unitlength}{3947sp}%
\begingroup\makeatletter\ifx\SetFigFont\undefined%
\gdef\SetFigFont#1#2#3#4#5{%
  \reset@font\fontsize{#1}{#2pt}%
  \fontfamily{#3}\fontseries{#4}\fontshape{#5}%
  \selectfont}%
\fi\endgroup%
\begin{picture}(5498,4640)(3068,-6043)
\put(8551,-5911){\makebox(0,0)[lb]{\smash{{\SetFigFont{20}{24.0}{\familydefault}{\mddefault}{\updefault}{\color[rgb]{0,0,0}$d_3$}%
}}}}
\put(5611,-2026){\makebox(0,0)[lb]{\smash{{\SetFigFont{20}{24.0}{\familydefault}{\mddefault}{\updefault}{\color[rgb]{0,0,0}$r_1$}%
}}}}
\put(5626,-5431){\makebox(0,0)[lb]{\smash{{\SetFigFont{20}{24.0}{\familydefault}{\mddefault}{\updefault}{\color[rgb]{0,0,0}$r_2$}%
}}}}
\put(3226,-4261){\makebox(0,0)[lb]{\smash{{\SetFigFont{20}{24.0}{\familydefault}{\mddefault}{\updefault}{\color[rgb]{0,0,0}$s$}%
}}}}
\put(8551,-3736){\makebox(0,0)[lb]{\smash{{\SetFigFont{20}{24.0}{\familydefault}{\mddefault}{\updefault}{\color[rgb]{0,0,0}$d_2$}%
}}}}
\put(8476,-1711){\makebox(0,0)[lb]{\smash{{\SetFigFont{20}{24.0}{\familydefault}{\mddefault}{\updefault}{\color[rgb]{0,0,0}$d_1$}%
}}}}
\end{picture}%